\documentclass[graybox]{svmult}

\usepackage{mathptmx}       
\usepackage{helvet}         
\usepackage{courier}        
\usepackage{type1cm}        
\usepackage{makeidx}         
\usepackage{graphicx}        
\usepackage{multicol}        
\usepackage[bottom]{footmisc}

\makeindex             

\newcommand{\vc}[1]{\textbf{\em #1}}

\begin{document}

\title*{Acceleration of Energetic Particles through Reconnection of Weakly Stochastic Magnetic Field}
\titlerunning{Acceleration via Turbulent Reconnection} 
\author{A. Lazarian$^1$, G. Kowal$^2$, B. Gouveia dal Pino$^2$ \& E. Vishniac$^3$}
\authorrunning{Lazarian et al.} 
\institute{$^1$ Astronomy Department, University of Wisconsin, Madison, WI 53706, US\\
$^2$ Instituto de Astronomia, Geof\'\i sica e Ci\^encias Atmosf\'ericas,
  Universidade de S\~ao Paulo, Rua do Mat\~ao, 1226 -- Cidade Universit\'{a}ria,
  CEP 05508-090, S\~ao Paulo/SP, Brazil\\
$^3$ Department of Physics and Astronomy, McMaster University, Hamilton, ON L8S, 4M1, Canada}
%
%
\maketitle

\abstract{Astrophysical media are turbulent and therefore reconnection should be treated in the presence
of pre-existing turbulence.  We consider the model of fast
magnetic reconnection in Lazarian \& Vishniac (1999) which predicts that the rate
of reconnection is controlled by the intensity and the injection scale of turbulent motions.
We provide new evidence of successful testing of the model and argue that the model
presents a generic set up for astrophysical reconnection events. 
We study particle acceleration that takes place in volumes of driven turbulence as well turbulent
volumes in the presence of large scale reconnection. 
We show that in the latter case the acceleration is of the first order Fermi type
thus supporting the model of acceleration proposed in Gouveia dal Pino \& Lazarian (2005).}


\section{Introduction}
\label{sec:1}

Astrophysical fluids are turbulent and this changes the physics of many processes. It has been long accepted that this dramatically alters
the nature of the cosmic ray propagation and acceleration (see Longair 2011), the research during the last decade also testifies that
the nature of the reconnection process is being radically changed (Lazarian \& Vishniac 1999, henceforth LV99, Kowal et al. 2009, 2012). Reconnection in
turbulent media was invoked by de Gouveia dal Pino \& Lazarian (2005, henceforth GL05, see also Lazarian 2005) in the model of the first order Fermi acceleration of cosmic rays
in reconnection sites. This paper discusses new testings of the LV99 model as well as new results on energetic particle acceleration in
the regions of reconnection of weakly turbulent magnetic field.   

\begin{figure}
  \includegraphics[width=.48\textwidth]{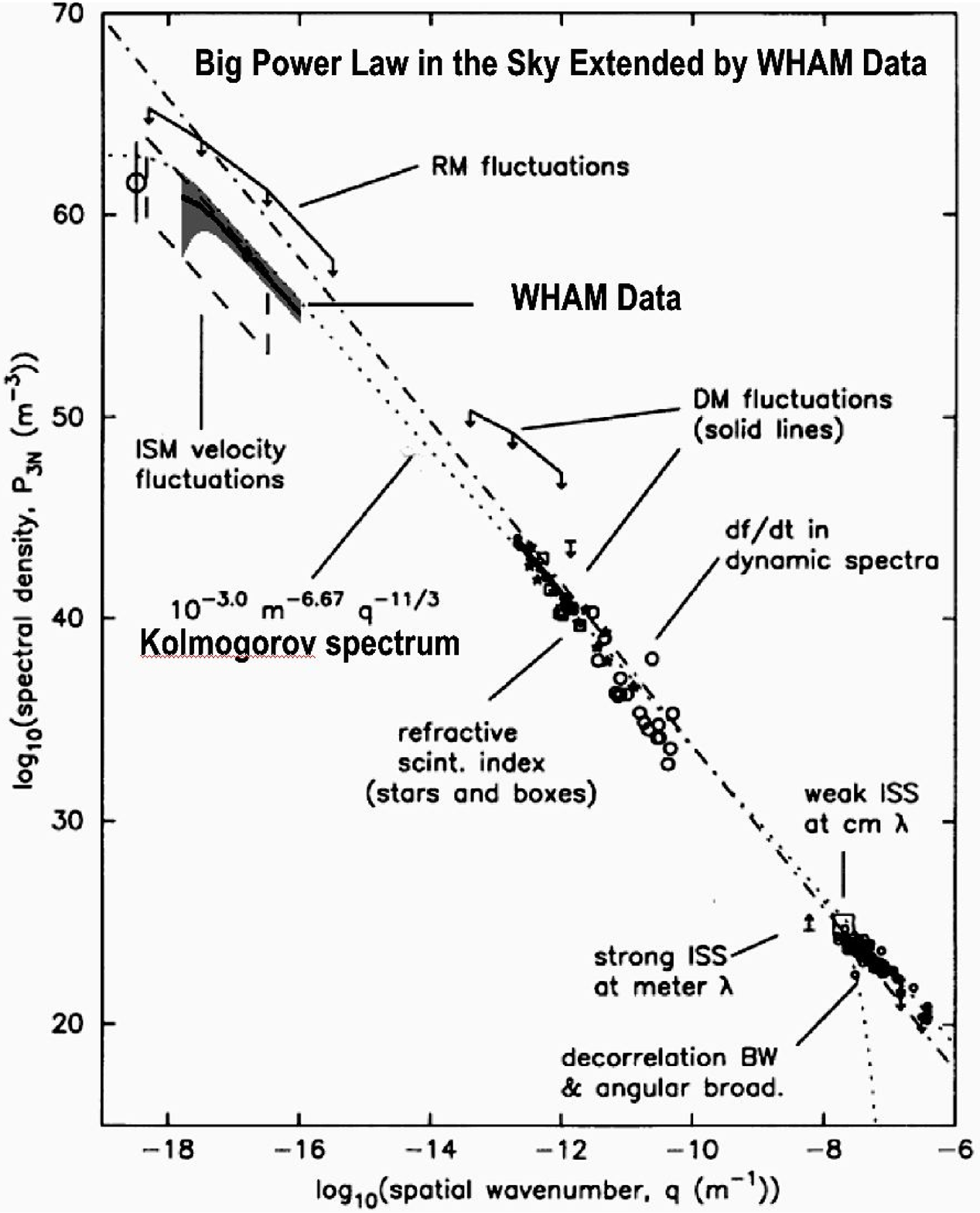}
   \includegraphics[width=.48\textwidth]{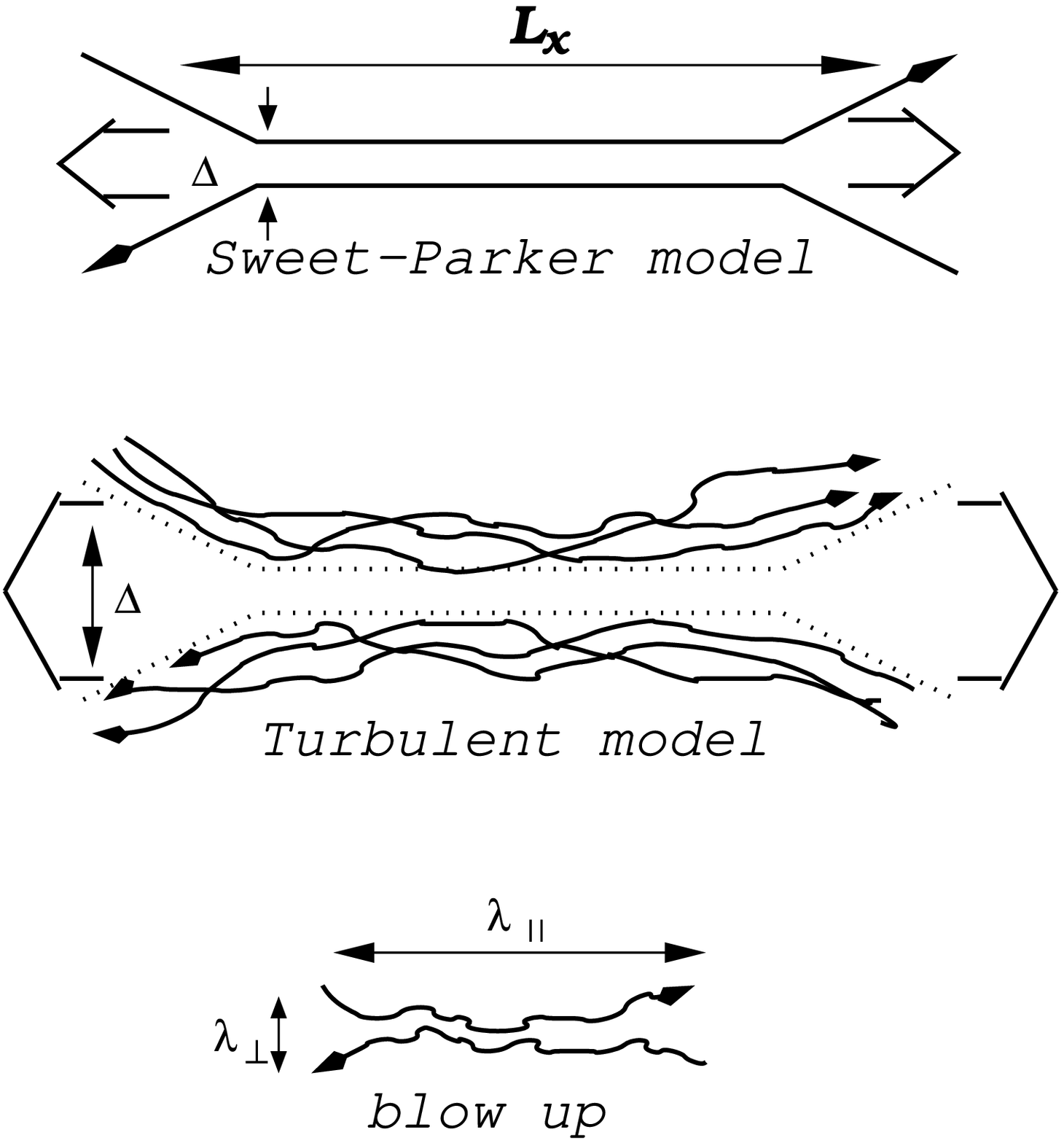}
  \caption{{\it Left panel}: Turbulence in the interstellar gas as revealed by electron density fluctuations. "Big Power Law in the Sky" in Armstrong et al. (1995) extended using WHAM data. The slope corresponds to that
of Kolmogorov turbulence. Modified from Chepurnov \& Lazarian (2010). {\it Right Panel}: Sweet-Parker model versus the model in LV99. Turbulence
makes the outflow region much wider and independent of resistivity. From Lazarian, Vishniac \& Cho 2004.}
 \label{CL}
\end{figure}

Astrophysical turbulence, in general, is {\it not} caused by reconnection. Therefore, while dealing with problems of reconnection and related particle acceleration it is proper to treat the turbulence as pre-existing. For instance, it has been known for decades that interstellar medium (ISM) is driven by violent supernovae explosions (McKee \& Ostriker 1977). By now it has been accepted that the ISM is turbulent on scales ranging from AUs to kpc (see Armstrong et al. 1995, Elmegreen \& Scalo 2004, Lazarian 2009). Figure \ref{CL} shows the turbulent power density plotted against the inverse of the scale length, with data at large scales, i.e. at small wavenumbers $q$ expanded using the Wisconsin H$_{\alpha}$ Mapper (WHAM) data on electron density fluctuations (Chepurnov \& Lazarian 2010). Another example is turbulence is solar wind (see
Leamon et al. 1998). 

The problem of magnetic reconnection is a long standing one. It is generally believed that a 
magnetic field embedded in a highly conductive
fluid preserves its topology for all time due to the magnetic fields being
frozen-in (see Alfven 1942).  At the same time, although ionized astrophysical objects,
like stars and galactic disks, are almost perfectly conducting, they show
indications of changes in topology, ``magnetic reconnection'', on dynamical time
scales (see Parker 1970).  Reconnection can be observed directly in the
solar corona (Yokoyama \& Shibata 1995), but can also be inferred from the
existence of large-scale dynamo activity inside stellar interiors (Parker 1993).  Solar flares are
usually associated with magnetic reconnection. More recent research shows that $\gamma$-ray busts can also have
a similar origin (Lazarian et al. 2003, Zhang \& Yan 2011)\footnote{LV99 model of reconnection predicts the
bursty character of reconnection when the initial state of magnetic field is close to the laminar one (see Lazarian \& Vishniac 2009).}. At the same time, a lot of previous work has concentrated on showing how
reconnection can be rapid in plasmas with very small collisional rates
(see Drake 2001, Drake et al. 2006). We feel that this substantially constrains astrophysical
applications of the corresponding reconnection models. 

Magnetic reconnection presents a known example of flux freezing violation. but it is conceivable that it was not taken seriously 
due to the unclear nature of fast reconnection (see Zweibel \& Yamada 2009 and references therein). 
Indeed, for years it was considered that fast reconnection required some special physical conditions and 
therefore "flux freezing" is fulfilled everywhere apart from some special zones.

The understanding of flux freezing in turbulent astrophysical environments has been challenged relatively recently and not
all the consequences of this radical change have been evaluated so far. LV99 identified magnetic field wandering, which is the inherent property of magnetized turbulent plasma, as the cause of fast, i.e. independent of resistivity, magnetic reconnection. They showed that in turbulent fluids magnetic fields should undergo constant reconnection and change their identity all the time. This implies that magnetic fields are not any more frozen into a perfectly conducting fluid if this fluid is turbulent as was explicitly stated first in Vishniac \& Lazarain (1999). Later, the challenge to the concept of "flux freezing" came from another side, i.e. from more formal mathematical studies of 
turbulent magnetic fields (see Eyink 2011a).  Eyink, Lazarian \& Vishniac (2011, henceforth ELV11) showed the consistency of these two approaches and established the equivalence of the LV99 treatment with that in more recent mathematical papers.

While the idea that turbulence can change the reconnection rates has been discussed in a number of earlier papers, the LV99 model was radically different from its predecessors. For instance, 
Mathaeus \& Lamkin (1985, 1986) performed 2D numerical
simulations of turbulence and provided arguments in favor of magnetic reconnection getting fast. However, the physics of the
processes that they considered was very different from that in LV99. For instance, the key process of field wandering of the LV99
model was been considered in Mathaeus \& Lamkin (1985, 1986). On the contrary, the components of their approach, e.g. X-point and
possible effects of heating and compressibility are not ingredients of the LV99 model. Other papers, e.g. \cite{speiser70} and \cite{jacobson84} explore the changes of the microscopic properties of the plasma induced by turbulence and consider how these changes can accelerate magnetic
reconnection. LV99 shows that the microscopic plasma properties are irrelevant for their model of reconnection (see testing in Kowal et al. (2009) and
more discussion of plasma effects in ELV11).

In what follows, we discuss the LV99 model in \S 2, present new numerical tests of the LV99 model in \S 3, demonstrate the efficiency
of energetic particle acceleration in turbulent reconnection regions in \S 4. In \S 5 we present astrophysical settings where we identified
the acceleration of particles in reconnection layers, in \S 6 and \S 7 we, respectively, provide the discussion and
summary of our results.

\section{Reconnection of Weakly turbulent magnetic field}
\label{sec:2}

To deal with strong, dynamically important magnetic fields LV99 proposed a model of 
fast reconnection in the presence of
sub-Alfv\'enic turbulence. It is important to stress that unlike laboratory
controlled settings, in astrophysical situations turbulence is preexisting,
arising usually from the processes different from reconnection itself. In fact,
any modeling of astrophysical reconnection should account for the turbulent state
of fluids and in most cases the turbulence does not arise from magnetic reconnection. The analogy here can
be as follows: turbulence that is experienced by the airplane does not arise from
the airplane motion, but preexist in the atmosphere.

LV99 identified stochastic wandering of the magnetic field-lines as the most
critical property of MHD turbulence which permits fast reconnection and obtained
analytical relations between the reconnection rate and the turbulence intensity
and the turbulence injection scale.

LV99 revealed a very intimate relation between turbulence and magnetic
reconnection and this connection was deepened by later research (ELV11, Lazarian, Eyink \& Vishniac 2012). 
First of all, LV99 showed that reconnection was a necessary ingredient
of MHD turbulence, this was the process that made the currently accepted picture
of MHD turbulence in Goldreich \& Sridhar (1995) self-consistent.  Moreover, further
research in ELV11 revealed that the expressions of
reconnection rate in LV99 can be obtained from the
concept of Richardson diffusion, which is the basic concept of fluid turbulence.  
\begin{figure}
\center
\includegraphics[width=0.48\textwidth]{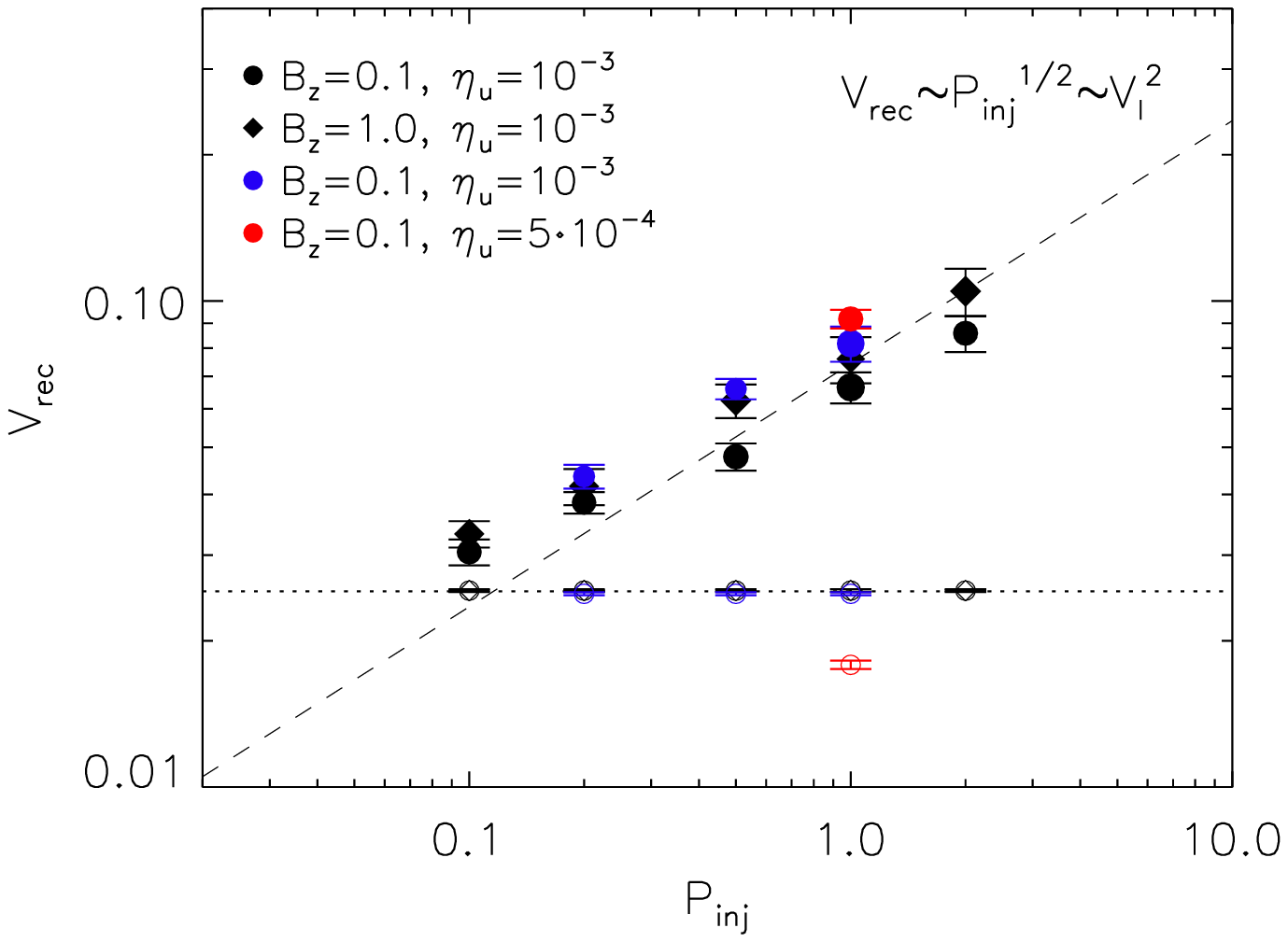}
\includegraphics[width=0.48\columnwidth]{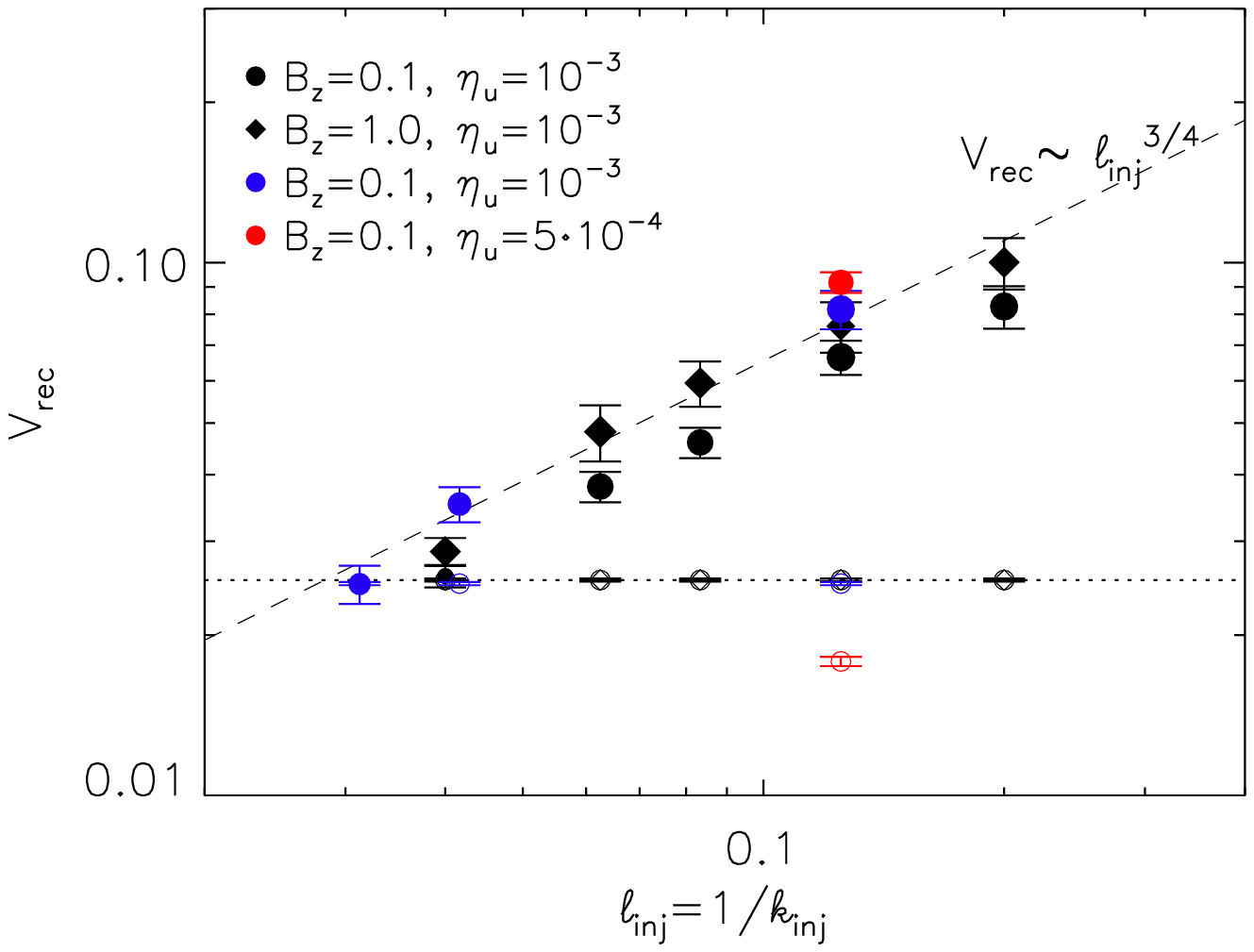}
\caption{ New and old testing of LV99 model. {\it Left Panel}: The dependence of the reconnection speed $V_{rec}$ on $P_{inj}$. Blue symbols show models with new driving in which the eddies where injected in magnetic field instead of velocity, as in the previous models (black symbols). A unique red symbol shows the reconnection rates from model with new driving in velocity performed with higher resolution (512x1024x512) and resistivity coefficient reduced to $\eta_{u}=5\cdot10^{-4}$.  {\it Right Panel}: The dependence of the reconnection speed $V_{rec}$ on $l_{inj}$ with additional models in which turbulence was driven in a new way as we describe in the text. 
Error bars represent the time variance of $V_{rec}$. The size of symbols corresponds to the error of $V_{rec}$. From Kowal et al. 2012.
\label{pow_dep}}
\end{figure}

\section{Numerical Testing of LV99 Model}

Testing of LV99 model of reconnection was performed in Kowal et al. (2009), (2012). Below we present some of the results obtained. 
In Figure~\ref{pow_dep} we see the results for varying amounts of input power, for fixed resistivity and injection scale as well as for the case of no turbulence at all.  The line drawn through the simulation points is for the LV99 predicted scaling, i.e. $V_{rec}\sim P^{1/2}$, where $P$ is the power of injected turbulence. We also see the results obtained with a new way of real space turbulence driving (see more in Kowal et al. 2012). This driving is
different from the Fourier space in Kowal et al. (2009). The results with both types of driving support LV99 model. In addition, Kowal et al. (2009) confirmed the predicted independence of the rate of reconnection on both Ohmic and anomalous resistivities, supporting LV99 conclusion that the reconnection is determined only by the properties of the turbulent flow.

\section{Reconnection and Particle Acceleration}

\begin{figure}[ht]
 \center
\includegraphics[width=0.48\textwidth]{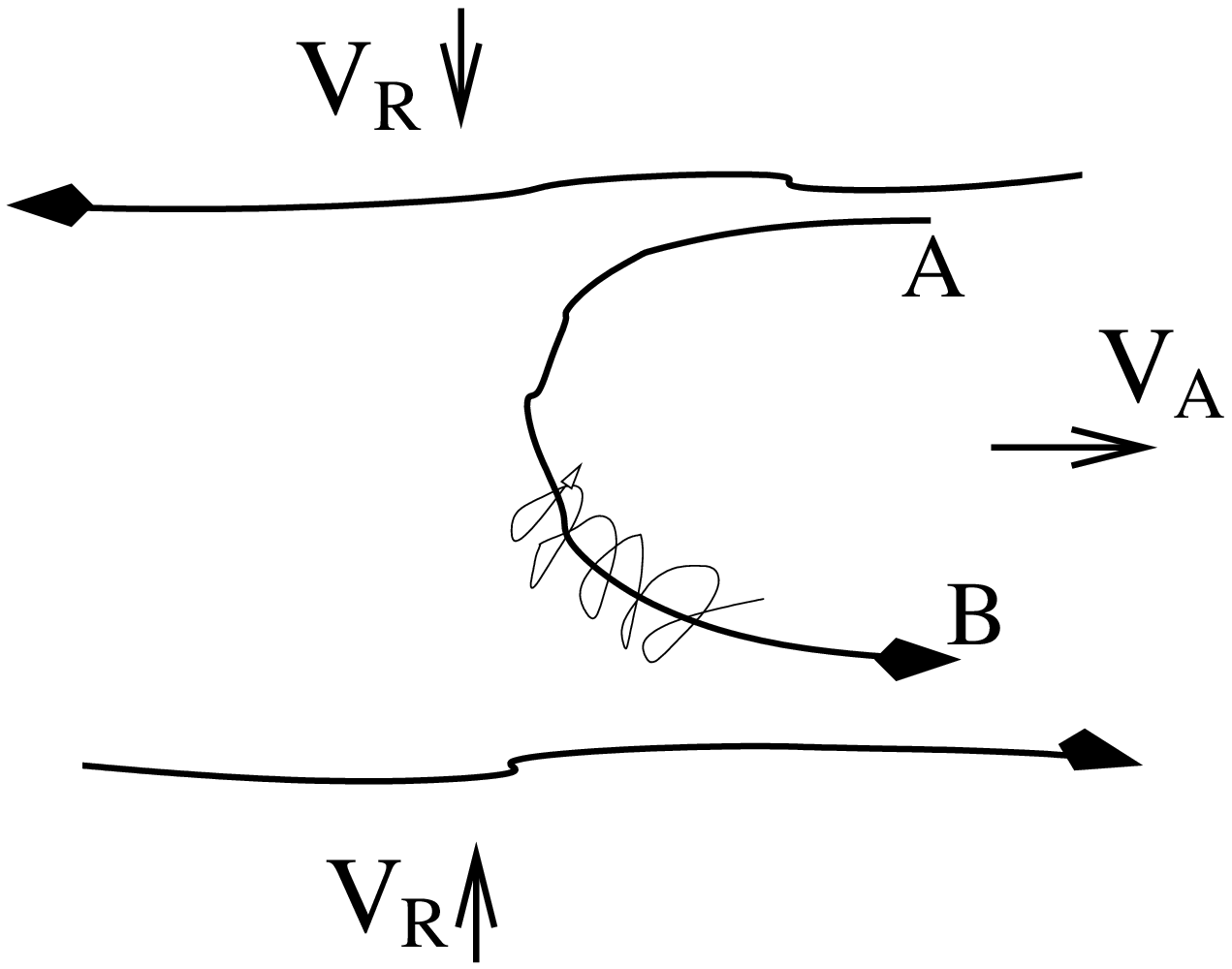}
 \includegraphics[width=0.48\textwidth]{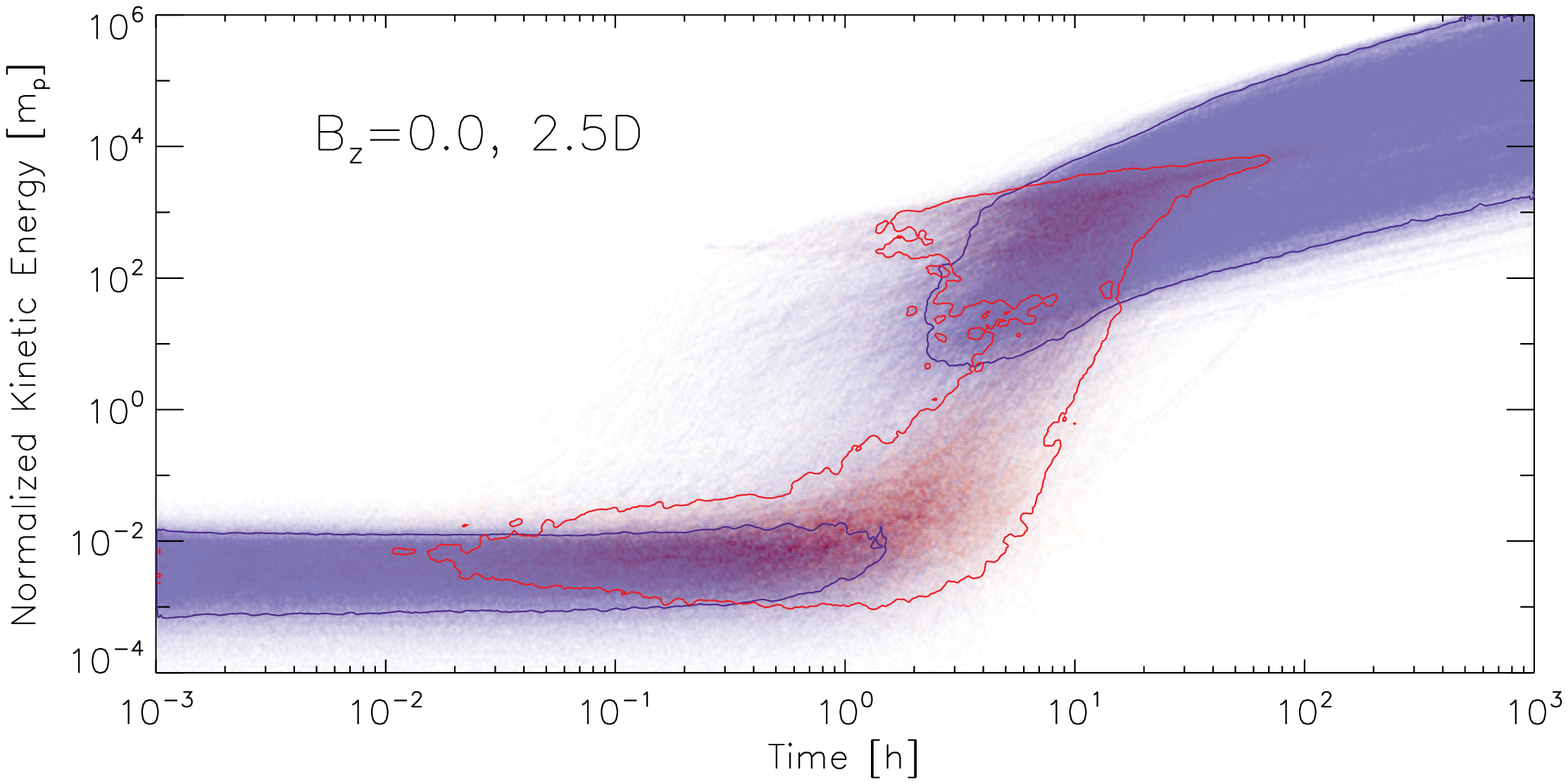}
 \includegraphics[width=0.48\textwidth]{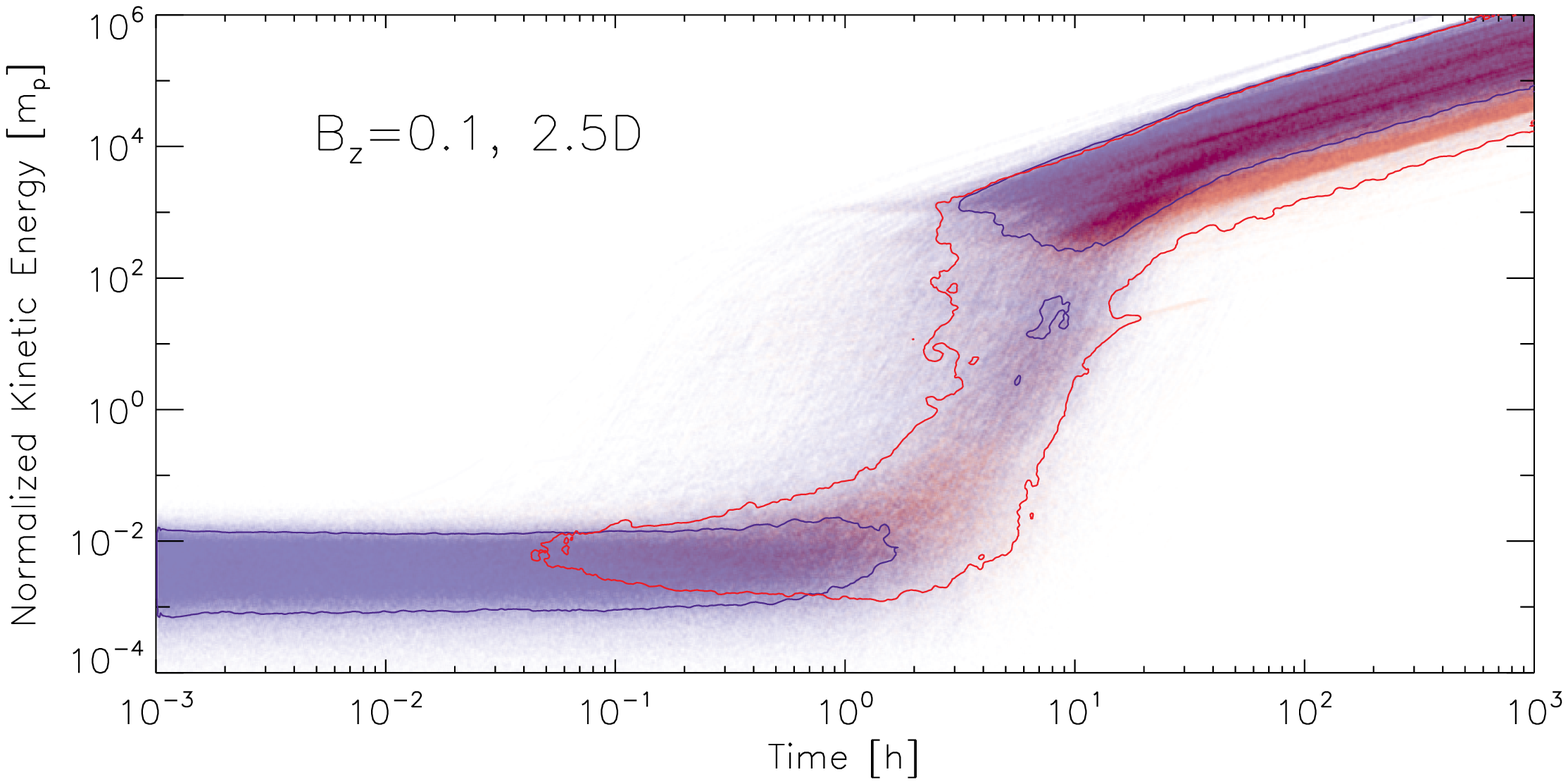}
 \includegraphics[width=0.48\textwidth]{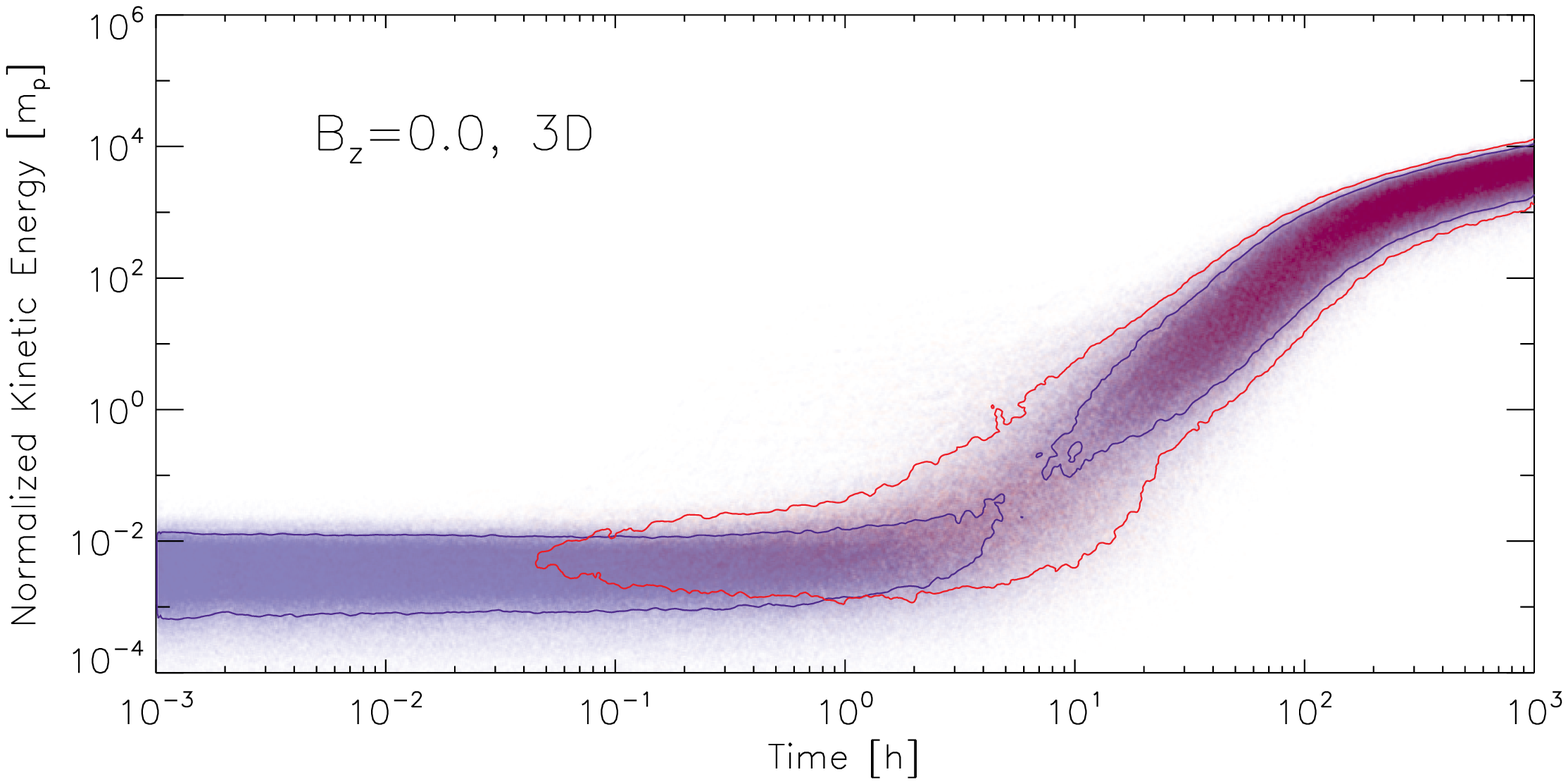}
 \caption{{\it Left upper corner}: Schematic of the first order Fermi acceleration
in the process of reconnecting turbulent magnetic field. From Lazarian 2005. 
{\it Other panels}. Numerical studies of particle acceleration. 
Kinetic energy evolution of a group of 10$^4$ protons in 2D models of
reconnection with a guide field $B_z$=0.0 and 0.1.
In the {\it right bottom panel} a fully 3D model with initial $B_z$=0.0 is presented.  The
colors show how the parallel (red) and perpendicular (blue) components of the
particle velocities increase with time. The contours correspond to values 0.1
and 0.6 of the maximum number of particles for the parallel and perpendicular
accelerations, respectively.  The energy is normalized by the rest proton mass
energy.  The background magnetized flow with multiple current sheet layers is at
time 4.0 in Alfv\'en time units. From Kowal et al. 2011
\label{accel}}
\end{figure}

\begin{figure*}
\center
\includegraphics[width=0.70\textwidth]{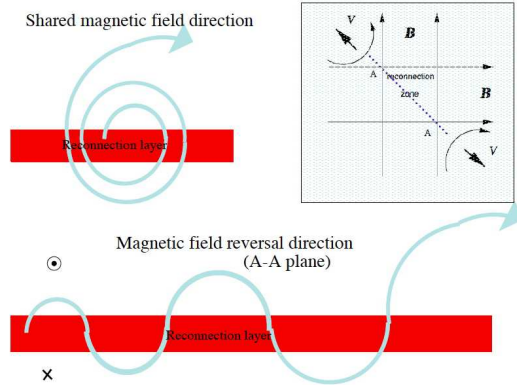}
\caption{Acceleration of energetic particles as the Larmor radius exceeds the thickness of current
sheet. The subpanel from LV99 shows the 3D structure of the reconnection region with the plane of
the magnetic field reversal (A-A) shown. In the shared magnetic field direction the energetic particle gains energy every period passing through contracting magnetic field. In the A-A plane the particle tend to escape if not prevented by magnetic bottles or scattering.
\label{perp}}
\end{figure*}

Magnetic reconnection results in shrinking of magnetic loops which induces the charged particles entrained on magnetic loops to get accelerated (see Figure \ref{accel}) as well as interaction with the bulk
magnetic mirrors of the flux (see Figure \ref{perp}). The processes discussed in GL05 (and also Lazarian 2005) for the LV99 reconnection and then the process depicted in Figure \ref{accel} was adopted for the collisionless reconnection in \cite{drake06}.  The physics of the acceleration is the same although GL05 appealed to the 3D magnetic bundles (see Figure \ref{accel} ), while \cite{drake06} considered 2D shrinking islands.  The latter is an artifact of the constrained 2D geometry. The difference in dimensions affects the acceleration efficiency according to \cite{kowal11}. GL05 claimed that the acceleration is of the first order Fermi type. This was confirmed in \cite{kowal12}. The calculations in 
GL05 of the spectrum were performed assuming that the particle bounces between mirrors that approach each other with the reconnection velocity. Below we descibe the numerical set up and the results of calculations.

In order to integrate the test particle trajectories we freeze in time a data
cube obtained from the MHD models of reconnection in
\cite{kowal09} and inject test thermal particles in the domain with random initial
positions and directions. For each
particle we solve the relativistic motion equation
\begin{equation} \frac{d}{d
t} \left( \gamma m \vc{u} \right) = q \left( \vc{E} + \vc{u} \times \vc{B} \right) , \label{eq:ptrajectory}
\end{equation}
where $m$, $q$ and $\vc{u}$ are the particle mass, electric charge and velocity,
respectively, $\vc{E}$ and $\vc{B}$ are the electric and magnetic fields,
respectively, $\gamma \equiv \left( 1 - u^2 / c^2 \right)^{-1}$ is the Lorentz
factor, and $c$ is the speed of light.  The electric field $\vc{E}$ is taken
from the MHD simulations $\vc{E} = - \vc{v} \times \vc{B} + \eta \vc{J}$, where
$\vc{v}$ is the plasma velocity, $\vc{J} \equiv \nabla \times \vc{B}$ is the
current density, and $\eta$ is the Ohmic resistivity coefficient.  We neglect
the resistive term above since its effect on particle acceleration is negligible
\cite{kowal11}.

In Figure~\ref{accel}, we present the time evolution of the kinetic
energy of the particles which have their parallel and perpendicular (red and
blue points, respectively) velocity components accelerated for three models of
reconnection.  The upper left panel shows the energy evolution for a 2D model
without the guide field.
Initially, the particles pre-accelerate by increasing their perpendicular
velocity component only.  Later we observe an exponential growth of energy
mostly due to the acceleration of the parallel component which stops after the
energy reaches values of 10$^3$--10$^4$~$m_p$ (where $m_p$ is the proton rest
mass energy).  Further on, particles accelerate their perpendicular component
only with smaller rate in a log-log diagram.  In 2.5D case, there is also
an initial slow acceleration of the perpendicular component followed by the
exponential acceleration of the parallel velocity component.  Due to the
presence of a weak guide field, the parallel component accelerates further to
higher energies at a similar rate as the perpendicular one.  This implies that
the presence of a guide field removes the restriction seen in the 2D model
without a guide field and allows the particles to increase their parallel
velocity components as they travel along the guide field, in open loops rather
than in confined 2D islands.  This result is reassured by the 3D model in the
bottom panel of Figure~\ref{accel}, where no guide field is necessary
as the MHD domain in fully three-dimensional.  In this case, we clearly see a
continuous increase of both components, which suggests that the particle
acceleration behavior changes significantly when 3D effects are considered,
where open loops replace the closed 2D reconnecting islands.

The observed increase in the parallel component of the particle momentum is obvious from looking in 
the cartoon in Lazarian (2005) presented in Figure \ref{accel}. Indeed, bouncing back and forth th particle gains parallel momentum. However, if the distance between the approaching magnetic mirrors considered in GL05 becomes comparable with the Larmor radius of an energetic particle, the particle undergoes regular 
acceleration in terms of its perpendicular momentum. This is illustrated in Fgure \ref{perp} for
the case of a particle having Larmor radius larger than the reconnection region thickness. Note,
that in LV99 model the reconnection region contains a microcosm of contracting and reconnection magnetic flux tubes. Thus the requirement on the Larmor radius are not particlular restrictive and
the increase of the perpendicular component of  will interact with the non-reconnected magnetic flux as it is shown in Figure \ref{perp}. Then the particle gains energy every period. However, unless one
 flux tube engalfs the flux of the opposite direction, particles must be returned back to the 
reconnection region by either magnetic bottles or scattering\footnote{It is important to mention that
too efficient scattering suppresses First order Fermi acceleration in incompressible fluids. Anisotropy of the particle momenta are required for the acceleration to be efficient.}. While the role of parallel and perpendicular First order Fermi acceleration as well as the effects of the back-reaction of particles on magnetic reconnection are discussed in more detail in our extended review in Lazarian et al. (2012), these issues remain exciting subjects to be studied in detail for years to come.  

\section{Reconnection and Acceleration: Astrophysical Examples}

Lazarian \& Opher (2009) identified the origin of anomalous cosmic rays measured by Voyagers as arising from magnetic reconnection
in heliosheath. As magnetic field lines of the magnetic spiral (arising from solar rotation) 
present in the solar wind get closer together magnetic reconnection is
expected. In view of solar wind being turbulent, the LV99 model of reconnection is relevant\footnote{Drake et al. (2010) followed
with a model appealing to the collisionless reconnection for which turbulence is not directly involved.}.

To explain the excess of the cosmic ray arrival from the direction towards heliotail Lazarain \& Desiati (2010) considered magnetic
reconnection in heliotail. The accumulation of magnetic flux of opposite polarities in heliotail arises from solar magnetic field 
reversals within 11 year cycle. Magnetic fields in heliotail are expected to be turbulent. The mechanism provides an attractive alternative to those discussed in the literature.

Magnetic reconnection in clusters of galaxies present an interesting way of accelerating energetic particles there (see Lazarian \& Brunetti 2011). 
The large scale reversals of magnetic fields in galaxy clusters are expected as a result of accretion flows, instabilities and galactic jets. 
Acceleration in reconnection regions may present an appealing alternative to the acceleration of cosmic rays by the intracluster magnetic turbulence (see Brunetti \& Lazarian 2007). 
Further work on the quantitative elaboration of the predictions of this and other astrophysical models that appeal to the acceleration of cosmic
rays in reconnection sites are necessary.

\section{Discussion}

First of all, we would like to stress that the LV99 model is not in conflict with the studies of magnetic reconnection in collisionless plasmas that have been a major thrust of the plasma physics community (see Shay et al. 1998, Daughton et al. 2006). Unlike latter studies, LV99 deals with turbulent environments. It is demonstrated in LV99, the local reconnection rates are influenced by plasma effects, e.g. kinetic effects of Hall effects, but the overall or global reconnection rate, i.e. the rate at which magnetic flux tubes reconnect, is determined by the turbulent broadening of the reconnection region. Thus, in the turbulent astrophysical media the rate of reconnection is not going to be affected by additional mechanisms.

Over the last decade, more traditional approaches to reconnection have 
changed considerably. At the time of its introduction, the models competing with LV99 
were modifications of the single X-point collisionless reconnection scheme first introduced by \cite{petschek64}. 
Those models had 
point-wise localized reconnection regions which were stabilized via plasma effects so that the
outflow opened up on larger scales. Such configurations would be difficult to 
sustain in the presence of random forcing,  which would be expected to collapse the reconnection layer. 
Moreover, \cite{CiaravellaRaymond08} argued that observations of solar flares were inconsistent with single X-point 
reconnection.

In response to these objections, more recent models of collisionless reconnection have acquired several 
features in common with the LV99 model. In particular, they have moved to consideration of volume-filling 
reconnection, (although it is not clear how this volume filling is achieved in the presence of a single 
reconnection layer (see Drake et al. 2006)).  While much of the discussion still centers around magnetic 
islands produced by reconnection, in three dimensions these islands are expected to evolve into 
contracting 3D loops or ropes due to tearing-type instabilities in electron-current layers 
(\cite{Daughtonetal08,Daughtonetal11}). This is broadly similar to what is depicted in Figure \ref{CL},  at least 
in the sense of introducing stochasticity to the reconnection zone. At the same time, although the 3D PIC simulation studies
are described as ``turbulent'', they do not exhibit the inertial-range power-law 
spectra observed in the magnetosphere and do not take into account either the pre-existing turbulence 
found in many of its regions (due to temperature anisotropy, velocity shear, Kelvin-Helmholtz instability, etc.)
or inertial-range turbulence generated as a consequence of reconnection itself  \cite{Zimbardoetal10}. 

The above remarks apply to the recent 3D PIC study Che et al. (2011), which 
observes micro-turbulence in the electron current layer during reconnection. The authors identify 
the source of this ``turbulence'' as a filamentation instability driven by current gradients, very
similar to a related instability in the Electron MHD (EMHD) model. The key aim of this work was to identify 
the term in the generalized Ohm's law which supplies the reconnection electric field  to break
the ``frozen-in'' condition. However, this study ignores the ambient inertial-range turbulence 
observed in the magnetosphere and other astrophysical plasmas, which may strongly modify 
laminar instabilities and reconnection process. 
 
We believe that in most astrophysical situations one has to deal with the {\it pre-existing turbulence}, 
which is the inevitable consequence of the high Reynolds number of astrophysical fluids 
and for which abundant empirical evidence exists. Such turbulence may modify or suppress instabilities, 
e.g. the tearing mode instability invoked in many studies \cite{Loureiroetal09, Bhattacharjeeetal09}.  
At the same time, LV99 model induces fast reconnection on dynamical time scales.

Our study shows that the acceleration of energetic particles in LV99-type reconnection layers
provides first order Fermi acceleration first described in de GL05. The
studies in Kowal et al. (2011, 2012) confirm the promise of this way of accelerating of particles in
various environments. They also pose questions of the relative importance of the parallel and
perpendicular accelerations and on the role of the backreaction of the accelerated particles.

\section{Summary}

The results of our studies can be very briefly summarized as follows:

1. Advances in the understanding of magnetic reconnection in the MHD regime, in
particular, related to the LV99 model of turbulent magnetic reconnection motivate
the studies of whether the reconnection in this regime can accelerate energetic
particles. New higher resolution testing of the LV99 model as well as simulations with
a different type of driving confirm that the reconnection is fast. 

2. Contracting magnetic loops in magnetic reconnection in 2D, in the MHD regime,
provide the acceleration analogous to that observed in PIC simulations, which
proves that the acceleration in reconnection regions is a universal process
which is not determined by the details of plasma physics. This confirms that the
process of acceleration in GL05 and Drake (2006) has the same nature.

3. Acceleration of energetic particles in 2D and 3D shows substantial
differences, which call for focusing on realistic 3D geometries of reconnection.
Our study also shows that the first order Fermi acceleration
dominates the second order Fermi that is also present in turbulent reconnection
layers.

\paragraph{Acknowledgements.}
Helpful comments of the anonymous referee improved the original manuscript. 
The research of AL is supported
by the Center for Magnetic Self-Organization in Laboratory and Astrophysical
Plasmas. AL also
acknowledged Humboldt Award at the Universities of Cologne and Bochum, as well
as the hospitality of the International Institute of Physics (Brazil) and the
Vilas Associate Award.  
GK and EMGDP acknowledge the support by the FAPESP grants no. 2006/50654-3 and
2009/50053-8, and the CNPq  grant no. 300083/94-7. This
research was also supported by the project TG-AST080005N through TeraGrid
resources provided by Texas Advanced Computing Center
(http://www.tacc.utexas.edu).

\end{document}